\documentclass[aps,prd,showpacs,twocolumn]{revtex4}
\usepackage{amssymb}
\usepackage{amsfonts}
\usepackage{amsmath}
\usepackage{graphicx}

\begin{document}

\title{Gupta--Bleuler's quantization of the anisotropic parity-even and
CPT-even electrodynamics of standard model extension}
\author{R. Casana, M. M. Ferreira Jr, F. E. P. dos Santos}
\affiliation{$^{a}${\small {Universidade Federal do Maranh\~{a}o (UFMA), Departamento de F%
\'{\i}sica, Campus Universit\'{a}rio do Bacanga, S\~{a}o Lu\'{\i}s - MA,
65080-805 - Brazil}}}

\begin{abstract}
We have established the Gupta-Bleuler quantization of the photon belonging
to the anisotropic parity-even sector of the CPT-even and Lorentz-violating
nonbirefringent electrodynamics of the standard model extension. We first
present a rule for the Maxwell electrodynamics to be successfully quantized
via Gupta-Bleuler technique in the Lorentz gauge. Recognizing the failure of
the Gupta-Bleuler method in the Lorentz gauge, $\partial _{\mu }A^{\mu }=0$,
for this massless LV theory, we argue that Gupta-Bleuler can be
satisfactorily implemented by choosing a modified Lorentz condition, $%
\partial _{\mu }A^{\mu }+\kappa ^{\mu \nu }\partial _{\mu }A_{\nu }=0$,
where $\kappa ^{\mu \nu }$ represents the Lorentz-violation in photon
sector. By using a plane-wave expansion for the gauge field, whose
polarization vectors are determined by solving an eigenvalue problem, and a
weak Gupta-Bleuler condition, we obtain a positive-energy Hamiltonian in
terms of annihilation and creation operators. The field commutation relation
is written in terms of modified Pauli-Jordan functions, revealing the
preservation of microcausality for sufficiently small LV parameters.
\end{abstract}

\maketitle

\section{Introduction}

The gauge sector of the standard model extension (SME) \cite{Colladay},
composed of a CPT-odd \cite{Jackiw} and a CPT-even part \cite{Mewes}, has
also been investigated in connection with Quantum Electrodynamics (QED) \cite%
{Soldati} and quantization aspects. The CPT-odd part was first addressed,
involving the analysis of consistency (causality, stability, and unitarity)
as well \cite{Adam,Baeta}. Similar consistency analysis about the CPT-even
sector, based on the evaluation of the Feynman propagator, was performed for
the parity-even and parity-odd sectors in Refs. \cite{Propag}. The
parity-odd nonbirefringent photonic sector was also investigated in Ref.
\cite{Schreck1}, which contains a broader analysis including the basis of
polarization vectors, coupling with fermions and evaluation of typical QED
processes. An analogue investigation was developed for the birefringent
CPT-even gauge sector of the SME \cite{Schreck2}. Moreover, to make possible
the treatment of infrared divergences that plague the modified QED by the
CPT-even gauge sector of the SME, a mass term for photons was supposed \cite%
{Massive photons}. Quantization and consistency aspects were also
investigated in the context of Lorentz-violating theories with higher
dimension operators \cite{Reyes}. The covariant quantization of the photonic
CPT-even sector was addressed in Ref. \cite{Hohensee}, where it was
discussed the quantization in the context of an indefinite metric Hilbert
space. It was also reported the necessity of using a weak Lorentz gauge
condition in order to correctly implement the Gupta-Bleuler procedure. The
covariant quantization of the CPT-even gauge sector of the SME was recently
discussed in Refs. \cite{Colladaycov,colladay2014gupta}, in the presence of
the Proca mass term in order to avoid some incompatibilities, turning the
quantization procedure free from contradictions. In these works, the photon
mass has really worked out as a regulator, yielding the definition of
polarization basis that {allows} the consistent quantization of the theory.

The present work is proposed to analyze the canonical quantization of the
Maxwell massless electrodynamics modified by the nonbirefringent
coefficients $\kappa ^{ij}$ contained in the full CPT-even tensor $%
(k_{F})^{\mu \nu \alpha \beta }$ of the SME. The key point is a modified
Lorentz gauge condition, including the\ LV tensor\ $\kappa ^{\mu \nu },$
which fulfills a general condition for implementing the covariant
quantization method for non-massive photons. A plane wave expansion is
written in terms of the creation and annihilation operators and polarization
vectors. The polarization vectors are achieved as the eigenvalues of an
operator defining the equation of motion in this gauge. The Hamiltonian
reveals to be positive-definite considering the smallness of the LV
parameters. The microcausality is investigated in terms of a generalized
Pauli-Jordan function which includes LV contributions. Once the noncovariant
quantization was performed, we discuss the Gupta-Bleuler quantization
procedure in this theory. Finally, we present our remarks and conclusions.

\section{The CPT-even electromagnetic sector of SME}

The CPT-even photonic sector of the SME described by the minimal Lagrangian
\begin{equation}
\mathcal{L}=-\frac{1}{4}F_{\mu \nu }F^{\mu \nu }-\frac{1}{4}\left(
k_{F}\right) ^{\mu \nu \alpha \beta }F_{\mu \nu }F_{\alpha \beta },
\label{SMELagrangian}
\end{equation}%
where the LV background $\left( k_{F}\right) ^{\mu \nu \alpha \beta }$\ has
the same symmetries as the Riemann's tensor and a null double trace, $\left(
k_{F}\right) ^{\mu \nu }{}_{\mu \nu }=0,$\ providing a total of 19
independent components, from which 10 are birefringent. The terms associated
to birefringence are strongly constrained by data of distant galaxies \cite%
{Mewes}, so we\ can retain only the remaining $9$\ nonbirefringent$,$\ which
are parameterized\ by a symmetric and traceless tensor $\kappa ^{\mu \nu }$\
\cite{mention} as
\begin{equation}
\left( k_{F}\right) _{\mu \nu \alpha \beta }=\frac{1}{2}\left( \eta _{\mu
\alpha }\kappa _{\nu \beta }-\eta _{\mu \beta }\kappa _{\nu \alpha }+\eta
_{\nu \beta }\kappa _{\mu \alpha }-\eta _{\nu \alpha }\kappa _{\mu \beta
}\right) ,  \label{1.4)}
\end{equation}%
where $\kappa ^{\nu \beta }=$ $\left( k_{F}\right) _{\mu }^{\text{ \ \ }\nu
\mu \beta }$. Such parameterization allows to write the Lagrangian density (%
\ref{SMELagrangian}) as
\begin{equation}
\mathcal{L}=-\frac{1}{4}F_{\mu \nu }F^{\mu \nu }-\frac{1}{2}\kappa _{\nu
\rho }F^{\mu \nu }F_{\mu }{}^{\rho }.  \label{NBLagrangian}
\end{equation}%
The equation of motion for the gauge field reads \
\begin{eqnarray}
0 &=&\left( \square +\kappa ^{\alpha \rho }{}\partial _{\alpha }\partial
_{\rho }\right) A^{\beta }+\square \kappa ^{\beta \rho }{}A_{\rho } \\
&&-\kappa ^{\beta \rho }{}\partial _{\rho }\partial _{\alpha }A^{\alpha
}-\partial ^{\beta }\left( \partial _{\alpha }A^{\alpha }+\kappa ^{\alpha
\rho }{}\partial _{\alpha }A_{\rho }\right) ,  \notag
\end{eqnarray}%
with the last term suggesting a generalized gauge fixing condition%
\begin{equation}
\partial _{\alpha }A^{\alpha }+\kappa ^{\alpha \rho }{}\partial _{\alpha
}A_{\rho }=0,  \label{LVCG}
\end{equation}%
in the place of the usual Lorentz condition. Thus, the equation of motion is
reduced as
\begin{equation}
\left( \square +\kappa ^{\alpha \rho }{}\partial _{\alpha }\partial _{\rho
}\right) A^{\beta }+\square \kappa ^{\beta \rho }{}A_{\rho }+\kappa ^{\beta
\sigma }{}\kappa ^{\alpha \rho }{}\partial _{\sigma }\partial _{\alpha
}A_{\rho }=0.
\end{equation}

The parity-even isotropic and anisotropic coefficients are $\kappa _{00}$
and $\kappa _{ij}$, respectively, while $\kappa _{0i}$ are the parity-odd
components. In the next sections we propose a correct implementation of the
Gupta-Bleuler quantization by using the gauge condition (\ref{LVCG})\ for
the anisotropic parity-even and CPT-even sector of the electrodynamics
described by the Lagrangian (\ref{NBLagrangian}).

\section{Gupta--Bleuler Quantization}

A covariant prescription is accomplished by the Gupta--Bleuler quantization
\cite{Gupta,Bleuler},\textbf{\ }which can become inconsistent in the context
of LV theories \cite{Colladaycov,colladay2014gupta}. This procedure
introduces in the Lagrangian density a covariance term breaking the local
gauge invariance but not eliminating any gauge field degree of freedom. For
instance, in the Maxwell electrodynamics, the Lorentz condition, $(\partial
_{\mu }A^{\mu })^{2}/2\xi $, works well in the Feynman gauge ($\xi =1$) \cite%
{NAKANISHI}.

The Lagrangian density for Maxwell electrodynamics in Lorentz gauge is%
\begin{equation}
\mathcal{L}=-\frac{1}{4}F^{\mu \nu }F_{\mu \nu }-\frac{1}{2\xi }\left(
\partial ^{\mu }A_{\mu }\right) ^{2},
\end{equation}%
which gives the following equation of motion, in momentum space, $O_{\mu \nu
}^{\left( M\right) }\tilde{A}^{\nu }\left( p\right) =0,$ where we have
defined the tensor
\begin{equation}
O_{\mu \nu }^{\left( M\right) }=p^{2}g_{\mu \nu }-\left( 1-\xi ^{-1}\right)
p_{\mu }p_{\nu },
\end{equation}%
whose determinant $-\xi ^{-1}\left( p^{2}\right) ^{4}$\ provides the
dispersion relation $p^{2}=0$. It is easy to verify that for $\xi \neq 1$,
the null-space of the matrix $\left. O_{\mu \nu }^{\left( M\right)
}\right\vert _{p^{2}=0}$ has dimension 3. On the other hand, for $\xi =1$,
the null-space of the matrix $\left. O_{\mu \nu }^{\left( M\right)
}\right\vert _{p^{2}=0}$\ has dimension 4. The idea following this
observation is that a set of four eigenvectors belonging to the nulls-pace
of the matrix $\left. O_{\mu \nu }^{\left( M\right) }\right\vert _{p^{2}=0}$%
, for $\xi =1$, constitute a set of polarization vectors allowing a plane
wave expansion for the gauge field. With a well defined set of polarization
vectors satisfying the dispersion relation, the implementation of the
Gupta-Bleuler quantization follows naturally.

We now use this rule to justify the failure of the Gupta-Bleuler method in
the context of the Lorentz-violating electrodynamics defined by Eq. (\ref%
{NBLagrangian}), for the Lorentz condition in the Feynman gauge,\ such as it
was recently reported \cite{Colladaycov,colladay2014gupta}.\ We start from
the Lagrangian density (\ref{NBLagrangian}) in the presence of the Lorentz
gauge term,
\begin{equation}
\mathcal{L}=-\frac{1}{4}F^{\mu \nu }F_{\mu \nu }-\frac{1}{2}\kappa _{\nu
\rho }F^{\mu \nu }F_{\mu }{}^{\rho }-\frac{1}{2\xi }\left( \partial ^{\mu
}A_{\mu }\right) ^{2},  \label{Lagrang2}
\end{equation}%
where $\kappa _{\nu \rho }$ is parameterized as
\begin{equation}
\kappa _{\mu \nu }=\ell \left( u_{\mu }v_{\nu }+u_{\nu }v_{\mu }\right) ,
\label{1.73)}
\end{equation}%
and $u_{\mu }$, $v_{\mu },$ are spacelike four-vectors, $u_{\mu }=\left( 0,%
\mathbf{u}\right) $, $v_{\mu }=\left( 0,\mathbf{v}\right) $, with $\mathbf{u}
$ and $\mathbf{v}$ being orthonormal vectors. In the Feynman gauge, $\xi =1$%
,\ the equation of motion of the gauge field reads $O_{\mu \nu }^{\left(
LF\right) }\left( p\right) \tilde{A}^{\nu }\left( p\right) =0$, where
\begin{eqnarray}
O_{\mu \nu }^{\left( LF\right) }\left( p\right) &=&\left( p^{2}+\kappa
_{\alpha \beta }p^{\alpha }p^{\beta }\right) g_{\mu \nu }+p^{2}\kappa _{\mu
\nu }  \label{XCXC} \\
&&-p_{\mu }\kappa _{\nu \beta }p^{\beta }-p_{\nu }\kappa _{\mu \beta
}p^{\beta }.  \notag
\end{eqnarray}%
The condition $\det O_{\mu \nu }^{\left( LF\right) }\left( p\right) =0$\
provides the three dispersion relations:\ $p^{2}=0$, and
\begin{align}
\boxplus \left( p\right) & =p^{2}+\kappa ^{\alpha \beta }p_{\alpha }p_{\beta
}=0,  \label{phys_1} \\[0.08in]
\boxtimes \left( p\right) & =\left( 1-\ell ^{2}\right) p^{2}+\kappa ^{\alpha
\beta }p_{\alpha }p_{\beta }+\left( \kappa ^{2}\right) ^{\alpha \beta
}p_{\alpha }p_{\beta }=0,  \label{phys_2}
\end{align}%
where $\left( \kappa ^{2}\right) _{\mu \nu }=\kappa _{\mu }{}^{\beta }\kappa
_{\beta \nu }$.\ The first one has\ multiplicity 2, being a nonphysical
dispersion relation, while the others have multiplicity 1 and are the
physical dispersion relations. It can be verified that the dimension of the
null space of $\left. \frac{{}}{{}}O_{\mu \nu }^{\left( LF\right)
}\right\vert _{p^{2}=0}$\ is 1, i. e., we have only one polarization vector
associated with\ the dispersion relation $p^{2}=0$. The same holds for the
other dispersion relations: the dimension of the null space of $\left. \frac{%
{}}{{}}O_{\mu \nu }^{\left( LF\right) }\right\vert _{\boxplus =0}$ and $%
\left. \frac{{}}{{}}O_{\mu \nu }^{\left( LF\right) }\right\vert _{\boxtimes
=0}$ are both 1. The null spaces of the tensor (\ref{XCXC}), evaluated in
all dispersion relations, provide a total of three eigenvectors. Hence, the
Gupta-Bleuler technique cannot be applied satisfactorily in this
Lorentz-violating electrodynamics, for the usual Lorentz condition in the
Feynman gauge.

We solve this problem by selecting the gauge condition (\ref{LVCG}) which
represents a Lorentz-violating generalization of the Lorentz condition. It
will reveal to be a suitable choice to successfully\ implement the
Gupta-Bleuler quantization in this model. Hence, the Lagrangian density (\ref%
{NBLagrangian}) becomes
\begin{equation}
\mathcal{L}=-\frac{1}{4}F^{\mu \nu }F_{\mu \nu }-\frac{1}{2}\kappa _{\nu
\rho }F^{\mu \nu }F_{\mu }{}^{\rho }\!-\!\frac{1}{2\xi }(\partial _{\mu
}A^{\mu }\!+\!\kappa ^{\mu \nu }\partial _{\mu }A_{\nu })^{2}.  \label{1.71)}
\end{equation}%
By working in Feynman gauge, $\xi =1$, the equation of motion in the
momentum space for the gauge field reads
\begin{equation}
\mathcal{O}_{\mu \nu }\left( p\right) \tilde{A}^{\nu }\left( p\right) =0,
\label{gaugeeqmov}
\end{equation}%
with $\mathcal{O}_{\mu \nu }$ is defined by
\begin{equation}
\mathcal{O}_{\mu \nu }\left( p\right) =g_{\mu \nu }\left( p^{2}+\kappa
^{\alpha \beta }p_{\alpha }p_{\beta }\right) +\kappa _{\mu \nu }p^{2}+\kappa
_{\mu \alpha }\kappa _{\beta \nu }p^{\alpha }p^{\beta },
\end{equation}%
whose determinant is $-\left[ \boxplus \left( p\right) \right] ^{3}\boxtimes
\left( p\right) $. Here, we also\ identify the two physical dispersion
relations given in Eqs. (\ref{phys_1}) and (\ref{phys_2}). \ The null space
of $\mathcal{O}_{\mu \nu }$, when the dispersion relation $\boxtimes \left(
p\right) =0$\ is satisfied, is 1\ providing one polarization vector. On the
other hand, the dimension of the null space of $\left. \frac{{}}{{}}\mathcal{%
O}_{\mu \nu }\right\vert _{\boxplus \left( p\right) =0}$\ is 3, i. e., there
are three polarization vectors associated with the dispersion relation $%
\boxplus \left( p\right) =0$, one of them becomes physical and the other two
are not. So, the gauge condition (\ref{LVCG}), with $\xi =1$, provides 4
polarizations vectors associated with the dispersion relations, fulfilling
the condition for implementing the Gupta-Bleuler quantization.

The canonical conjugate momentum is%
\begin{equation}
\pi ^{\mu }=-\left( g^{\mu \nu }+\kappa ^{\mu \nu }\right) \dot{A}_{\nu },
\label{1.78)}
\end{equation}%
which allows to impose the following canonical commutation relations
\begin{equation}
\left[ A_{\mu }\left( t,\mathbf{x}\right) ,\pi ^{\nu }\left( t,\mathbf{y}%
\right) \right] =i\delta _{\mu }{}^{\nu }\delta ^{3}\left( \mathbf{x}-%
\mathbf{y}\right) ,  \label{STANDARDGB1}
\end{equation}%
whereas the other are null.

The modified Lorentz condition, $\partial _{\mu }\left( A^{\mu }+\kappa
^{\mu \nu }A_{\nu }\right) =0$, allows to define four polarization vectors
which can be used to propose a solution for Eq. (\ref{gaugeeqmov}) as a
plane-wave expansion,
\begin{equation}
A_{\mu }\left( x\right) =\!\sum_{\lambda =0}^{3}\!\int \!\widehat{{d^{3}%
\mathbf{p}}}^{(\lambda )}\left[ a_{\left( \lambda \right) }\left( \mathbf{p}%
\right) e^{-ix\cdot p^{\left( \lambda \right) }}+h.c.\right] \varepsilon
_{\mu }^{\left( \lambda \right) }\left( \mathbf{p}\right) ,  \label{planeWGB}
\end{equation}%
where $\widehat{{d^{3}\mathbf{p}}}^{(\lambda )}={d^{3}\mathbf{p/}}\sqrt{%
\left( 2\pi \right) ^{3}2E^{\left( \lambda \right) }}$, the annihilation and
creation operators are described by $a_{\left( \lambda \right) }\left(
\mathbf{p}\right) $ and $a_{\left( \lambda \right) }^{\dagger }\left(
\mathbf{p}\right) $, respectively. The polarization vectors $\varepsilon
_{\mu }^{\left( \lambda \right) }\left( \mathbf{p}\right) $ satisfy the
following eigenvalue equation:%
\begin{equation}
\mathcal{O}^{\mu \nu }\varepsilon _{\nu }^{\left( \lambda \right) }=\alpha
^{\left( \lambda \right) }\left( g^{\mu \nu }+\kappa ^{\mu \nu }\right)
\varepsilon _{\nu }^{\left( \lambda \right) }.  \label{1.87)}
\end{equation}%
The eigenvalues\ $\alpha ^{\left( \lambda \right) }$\ are%
\begin{align}
\alpha ^{\left( 0\right) }& =\alpha ^{\left( 1\right) }=\alpha ^{\left(
2\right) }=p^{2}+\kappa _{\mu \nu }p^{\mu }p^{\nu }, \\[0.3cm]
\alpha ^{\left( 3\right) }& =p^{2}+\frac{\kappa _{\mu \nu }p^{\mu }p^{\nu
}+\left( \kappa ^{2}\right) _{\mu \nu }p^{\mu }p^{\nu }}{1-\ell ^{2}}.
\end{align}%
yielding the four polarization vectors
\begin{align}
\varepsilon _{\mu }^{\left( 0\right) }& =\left( 1,\mathbf{0}\right)
,~\varepsilon _{\mu }^{\left( 1\right) }=\left( 0,{\varepsilon }_{i}^{\left(
1\right) }\right) , \\
\varepsilon _{\mu }^{\left( 2\right) }& =\left( 0,{\varepsilon _{i}^{\left(
2\right) }}\right) ~,~\ \varepsilon _{\mu }^{\left( 3\right) }=\left( 0,{%
\varepsilon }_{i}^{\left( 3\right) }\right) ,  \label{1.83a}
\end{align}%
with%
\begin{align}
{\varepsilon }_{i}^{\left( 1\right) }& =n^{\left( 1\right) }p_{i}~,~{%
\varepsilon }_{i}^{\left( 3\right) }=n^{\left( 3\right) }\left(
d^{-1}_{\left( 2\right) }\right) _{ij}\epsilon _{jka}p_{k}w_{a},
\label{tribase_1} \\[0.2cm]
{\varepsilon _{i}^{\left( 2\right) }}& =n^{\left( 2\right) }\left[ \left(
d_{\left( 2\right) }\right) _{ab}p_{a}p_{b}w_{i}-\left( p_{a}w_{a}\right)
p_{i}\right] ,  \label{tribase_3}
\end{align}%
where\ $w_{i}=\epsilon _{ijk}u_{j}v_{k}$, and $n^{\left( i\right) }$\ are
normalization constants.\ The polarization vectors satisfy the normalization%
\textbf{\ }condition given by $\varepsilon _{\mu }^{\left( \lambda \right)
}\left( g^{\mu \nu }+\kappa ^{\mu \nu }\right) \varepsilon _{\nu }^{\left(
\lambda ^{\prime }\right) }=g^{\lambda \lambda ^{\prime }},$ while the
completeness relation reads
\begin{equation}
\sum_{\lambda =0}^{3}g^{\lambda \lambda }\varepsilon _{\mu }^{\left( \lambda
\right) }\varepsilon _{\nu }^{\left( \lambda \right) }=\left( g+\kappa
\right) _{\mu \nu }^{-1}.  \label{completnessGB}
\end{equation}

The energy for each polarization vector is
\begin{align}
E^{\left( 0\right) }& =E^{\left( 1\right) }=E^{\left( 2\right) }=\left\vert
\mathbf{p}\right\vert \sqrt{1-\kappa _{ij}p_{i}p_{j}/\mathbf{p}^{2}}, \\%
[0.2cm]
E^{\left( 3\right) }& =\left\vert \mathbf{p}\right\vert \sqrt{1-\frac{\kappa
_{ij}p_{i}p_{j}+\left( \kappa ^{2}\right) _{ij}p_{i}p_{j}}{\left( 1-\ell
^{2}\right) \mathbf{p}^{2}}},
\end{align}%
which coincides with the dispersion\ relations obtained in Ref. \cite{Propag}%
. The energies are real numbers whenever the Lorentz-violating parameters $%
\kappa _{ij}$ are sufficiently small. In order to satisfy the canonical
commutation relation (\ref{STANDARDGB1}), the creation and annihilation
operators must satisfy the standard commutation relations,
\begin{equation}
\left[ a_{\left( \lambda \right) }^{\dag }\left( p\right) ,a_{\left( \lambda
^{\prime }\right) }\left( q\right) \right] =g_{\lambda \lambda ^{\prime
}}\delta ^{3}\left( \mathbf{p}-\mathbf{q}\right) ,  \label{STANDARDGBC}
\end{equation}%
and the others being null. The goal of our choice for the polarization (\ref%
{1.83a})\ is to express the quantum Hamiltonian as an explicit sum of the
contributions of each polarization mode, as required,
\begin{equation}
H=-\sum_{\lambda =0}^{3}\int d^{3}\mathbf{p}~g_{\lambda \lambda }E^{\left(
\lambda \right) }N^{\left( \lambda \right) },  \label{hamiltoniaGB}
\end{equation}%
where $N^{\left( \lambda \right) }=a_{\left( \lambda \right) }^{\dag
}a_{\left( \lambda \right) }$ is the number operator counting $\left(
\lambda \right) $ mode. Despite the Hamiltonian can be expressed in a simple
form, it is not positive-definite. Moreover, at operator level, the gauge
condition $\partial _{\mu }A^{\mu }+\kappa ^{\mu \nu }\partial _{\mu }A_{\nu
}=0$ is not compatible with the commutation relation (\ref{STANDARDGB1}).
From Eq. (\ref{STANDARDGBC}), we observe that the temporal creation and
annihilation operators, $a_{\left( 0\right) }^{\dag }\ $and $a_{\left(
0\right) }$, satisfy a commutation relation with changed signal, providing
states with negative norm. These problems can be solved in a covariant way\
by imposing the Gupta-Bleuler condition\cite{Gupta,Bleuler}, i.e., the
physical states $\left\vert \varphi \right\rangle $ are those providing\
null expectation value\ for the modified gauge condition,
\begin{equation}
{\left\langle \varphi \right\vert (g^{\mu \nu }+\kappa ^{\mu \nu })\partial
_{\mu }A_{\nu }\left\vert \varphi \right\rangle }=0.  \label{STRONG}
\end{equation}%
This is a strong operator condition. The physical states can be selected
imposing a weaker operator condition,%
\begin{equation}
(g^{\mu \nu }+\kappa ^{\mu \nu })\partial _{\mu }A_{\nu }^{\left( +\right) }{%
\left\vert \varphi \right\rangle }=0,  \label{weak}
\end{equation}%
where the gauge field was decomposed in positive and negative frequencies, $%
A_{\mu }=A_{\mu }^{\left( +\right) }+A_{\mu }^{\left( -\right) }$,
respectively. We now explicitly implement the weak condition (\ref{weak}) in
the plane-wave expansion (\ref{planeWGB}) of the gauge field, attaining
\begin{equation}
\sum_{\lambda =0}^{3}\int \!\widehat{{d^{3}\mathbf{p}}}^{(\lambda
)}e^{-ix\cdot p^{\left( \lambda \right) }}\left[ p_{\mu }^{\left( \lambda
\right) }\left( g^{\mu \nu }+\kappa ^{\mu \nu }\right) \varepsilon _{\nu
}^{\left( \lambda \right) }\right] a_{\left( \lambda \right) }{\left\vert
\varphi \right\rangle }=0.  \label{GAUGEPLANEW}
\end{equation}%
It is verified that contributions coming from polarizations $\varepsilon
_{\nu }^{\left( 2\right) }$\ and $\varepsilon _{\nu }^{\left( 3\right) }$\
are null. The remaining polarizations yield $p_{\mu }^{\left( \lambda
\right) }\left( g^{\mu \nu }+\kappa ^{\mu \nu }\right) \varepsilon _{\nu
}^{\left( \lambda \right) }=\left( -1\right) ^{\lambda }E^{\left( \lambda
\right) },$ for $\lambda =0,1$, with $E^{\left( 1\right) }=E^{\left(
0\right) }$, which allows to achieve the following constraint for the
physical states:
\begin{equation}
\left[ a_{\left( 0\right) }-a_{\left( 1\right) }\right] {\left\vert \varphi
\right\rangle }=0.  \label{1.99)}
\end{equation}%
The last expression links the expectation value of the scalar and
longitudinal photon operator numbers,
\begin{equation}
{\left\langle \varphi \right\vert }N^{\left( 0\right) }{\left\vert \varphi
\right\rangle }={\left\langle \varphi \right\vert }N^{\left( 1\right) }{%
\left\vert \varphi \right\rangle ,}  \label{GBcondiction}
\end{equation}%
a supplementary condition (\ref{GBcondiction}) which solves the problem
concerning the negative energy contributions in the Hamiltonian (\ref%
{hamiltoniaGB}). Once we have successfully quantized this Lorentz-violating
electrodynamics, we can also compute the covariant gauge field commutation
relation
\begin{eqnarray}
\left[ A_{\mu }\left( x\right) ,A_{\nu }\left( y\right) \right] &=&\int
\sum_{\lambda =0}^{3}\frac{d^{3}\mathbf{p}}{\left( 2\pi \right)
^{3}2E^{\left( \lambda \right) }}  \label{ccallT} \\
&&\hspace{-1cm}\times \left( e^{i\left( x-y\right) \cdot p^{\left( \lambda
\right) }}-e^{-i\left( x-y\right) \cdot p^{\left( \lambda \right) }}\right)
g_{\lambda \lambda }T_{\mu \nu }^{\left( \lambda \right) },  \notag
\end{eqnarray}%
where $T_{\mu \nu }^{(\lambda )}=\varepsilon _{\mu }^{(\lambda
)}(p)\,\varepsilon _{\nu }^{(\lambda )}(p)$. Using the dispersion relations
and the completeness relation (\ref{completnessGB}), the result is
\begin{eqnarray}
\left[ A_{\mu }(x),A_{\nu }(y)\right] &=&T_{\mu \nu }^{(3)}(i\partial
)i\Delta ^{(3)}(x-y)  \label{1.105)} \\[0.2cm]
&&\hspace{-0.5cm}-\left[ (g+\kappa )_{\mu \nu }^{-1}+T_{\mu \nu
}^{(3)}(i\partial )\right] i\Delta ^{(2)}(x-y),  \notag
\end{eqnarray}%
with $\Delta ^{\left( 2\right) }\left( x-y\right) $ and $\Delta ^{\left(
3\right) }\left( x-y\right) $ being the generalized Pauli Jordan functions
defined by
\begin{equation}
\Delta ^{\left( \beta \right) }\left( x_{0},\mathbf{x}\right) =-\frac{%
\varepsilon \left( x_{0}\right) \delta \left( \left( x_{0}\right)
^{2}-\left( d_{\left( \beta \right) }^{-1}\right) _{ij}x_{i}x_{j}\right) }{%
2\pi \sqrt{\det \left( d_{\left( \beta \right) }\right) }},
\label{modifedPJ}
\end{equation}%
with%
\begin{equation}
\left( d_{\left( 2\right) }\right) _{ij}=\delta _{ij}-\kappa _{ij}~,~\ \
\left( d_{\left( 3\right) }^{-1}\right) _{ij}=\delta _{ij}+\kappa _{ij}.
\end{equation}

The argument of the $\delta-$function in (\ref{modifedPJ}) preserves the
light-cone structure (deformed or not) whenever\textbf{\ }$\left( d_{\left(
\beta \right) }^{-1}\right) _{ij}x_{i}x_{j}>0$. Such requirement is
guaranteed if the matrix $\left( d_{\left( \beta \right) }^{-1}\right) $ is
positive-definite, which holds since the Lorentz-violating parameters $%
\kappa _{ij}$ are sufficiently small (the same condition that yields energy
positivity). This way, the function $\Delta ^{\left( \beta \right) }\left(
x\right) $ vanishes for spacelike vectors implying microcausality is assured.

\section{Final remarks}

We conclude that we have successfully implemented the Gupta-Bleuler method
for this anisotropic parity-even and CPT-even electrodynamics
withoutrecourse to an infrared regularization or to a small mass for the
photon field, as already known in the literature. Our results can also be
verified via the Dirac's quantization method for constrained systems. This
formalism shows that in this LV electrodynamics, the temporal gauge $%
(A_{0}=0)$ is not compatible with the usual Coulomb gauge $(\partial
_{j}A_{j}=0),$ but with a modified LV version, $(\delta _{ij}-\kappa
_{ij})\partial _{i}A_{j}=0$, as it will reported elsewhere \cite%
{papercompleto}.

We believe that the covariant quantization formalism here derived may be
applied to other coefficients of the CPT-even tensor, as the parity-odd
ones, as far as in the context of other electrodynamic models endowed with
anisotropic or higher dimension terms.

\begin{acknowledgments}
The authors thank to CNPq, CAPES and FAPEMA (Brazilian research agencies)
for financial support.
\end{acknowledgments}

\end{document}